\documentstyle[12pt,aaspp4,lscape,flushrt]{article}

\def\be{\begin{equation}}
\def\ee{\end{equation}}
\def\kms{km~s$^{-1}$}
\def\cm2{cm$^{-2}$}

\def\lya{{\rm Ly}$\alpha$}
\def\lyb{Ly$\beta$}

\newcommand{\lsim}{\ \raise -2.truept\hbox{\rlap{\hbox{$\sim$}}\raise5.truept
        \hbox{$<$}\ }}
\newcommand{\gsim}{\ \raise -2.truept\hbox{\rlap{\hbox{$\sim$}}\raise5.truept
        \hbox{$>$}\ }} 

\begin{document} 

\input epsf 

\title{
The Lyman--$\alpha$ forest of a Lyman Break Galaxy: VLT Spectra 
of MS1512--cB58 at $z=2.724$\footnote{Based on observations collected at
the European Southern Observatory, the VLT/Kueyen telescope, ESO,
Paranal, Chile (ESO Programme 65.O-0471).}}

\author{S. Savaglio\footnote{Present address: The Johns Hopkins
University, 3400 North Charles Street, Baltimore,
MD21218, USA}}
\affil{\it Osservatorio Astronomico di Roma, Via di
Frascati 33, I--00040 Monteporzio, Italy}

\author{N. Panagia and P. Padovani\footnote{On leave of absence from
Dipartimento di Fisica, II Universit\`a di Roma "Tor Vergata", Italy}}

\affil{\it Space Telescope Science Institute, 3700 San Martin Drive, 
        Baltimore, MD21218, USA}

\affil{\it On assignment from the Space Science Department of the European
Space Agency}

\begin{abstract} 

The high redshift galaxy MS1512--cB58 ($z=2.724$, $m_V=20.64$) has
been observed with the very efficient high resolution echelle
spectrograph VLT/UVES. Although this is a very challenging
observational program for a Southern hemisphere telescope (the galaxy
is located at $+36^\circ$ declination), high resolution spectra (FWHM
$\simeq 26$ \kms) have revealed, with unprecedented detail along a
galaxy sight line, the Lyman--$\alpha$ forest due to intervening
clouds in the intergalactic medium (IGM). The mean depression $D_A$
due to IGM absorption blueward of the galaxy \lya~wavelength and the
number density $dn/dz$ of \lya~clouds have been compared with
equivalent results obtained for QSO sight lines at similar
redshifts. Our results indicate a possible excess of absorption close
to the galaxy.  The mean depression at $\sim 150~h_{65}^{-1}$ Mpc
comoving ($\Omega_m=0.3$, $\Omega_\Lambda=0.7$) from the galaxy is
$D_A=0.36\pm0.03$, to be compared with $0.22\pm0.04$, expected from a
best fit to QSO sight lines. In the same region ($z\simeq2.610$), the
number density of lines with HI column density in excess of $10^{14}$
atoms \cm2 is also $\sim 3\sigma$ larger than expected.  This high
density region is at least 60 $h_{65}^{-1}$ Mpc comoving wide, but the
large \lya~absorption of the galaxy itself prevents us from detecting a
possible  structure extending down to the galaxy.

This excess of \lya~clouds is suggestive of two possible scenarios.
One is the presence of a super cluster of \lya~clouds not associated
with cB58. The other is a high density of gas associated with the
environment of cB58. Indeed, a hint of the complexity of cB58 and
possibly its environment is given by the huge velocity range (almost
1000 \kms) between the optical emission of star forming regions and UV
absorption of its interstellar medium.

\end{abstract}


\section{INTRODUCTION}

Lyman break galaxies (LBGs) are by definition characterized by a
strong absorption longward of the \lya~wavelength due to intervening
intergalactic gas and intrinsic UV emission of strong star formation, and
so can easily be found using the color--color technique (Steidel et
al., 1996). This technique is efficient for redshifts larger than 2,
where intergalactic absorption is more pronounced and the galaxy UV
flux is redshifted to the optical, allowing observations from the ground. 

The rest frame UV spectrum of a galaxy bears the signature of its
interstellar medium through absorption lines and stellar activity
through line and broad band emission.  Although low resolution UV
spectra of galaxies are available for the local universe, high
resolution data are strongly limited by the low efficiency of space
telescopes. Only recently the far UV spectrograph (FUSE) started to
deliver high resolution (HR, $R\sim10,000$) spectra of nearby star
forming galaxies, revealing for the first time complex structures
(Heckman et al., 2001).
 
In principle, high redshift galaxies can benefit of the fact that
their UV emission is redshifted to the optical and can, therefore, be
studied by large ground-based telescopes. However, their apparent
optical magnitude is generally fainter than $\sim 23$ and HR
spectroscopy cannot be performed. Other possible candidates for HR
spectroscopy are the very powerful radio galaxies, but no radio galaxy
at redshift larger than $z=1.5$ and brighter than $m\sim 21$ has been
found so far. This limit is just a little beyond the HR capabilities
of the largest telescopes.

The chance to observe high redshift galaxies at high spectral
resolution is then confined to gravitational lensed galaxies and
gamma ray burst (GRB) afterglows. The latter are indeed potentially
very interesting for future observations (Fiore et al., 2000) thanks
to the advent of new X--ray satellites (HETE2 and in the near future
SWIFT) that are capable of prompt identification of GRBs.  The former
are still penalized by a limited number of sufficiently bright
targets.  A unique case is the galaxy MS1512--cB58 ($z=2.724$).

This very bright gravitational lens LBG ($m_V=20.65$, Ellingson et
al., 1996) is within the state of the art capabilities of HR
spectrographs of the largest ground based telescopes. It was
serendipitously discovered by Yee et al. (1996) during the CNOC
cluster redshift survey and is gravitationally lensed by the
cluster MS1512+36 at $z=0.37$ (Seizt et
al. 1998). The gravitational magnification has been studied in detail
by Seitz et al. (1998) who found that the $2\times 0.2$ arcsec$^2$
arclet brightness is amplified $\sim 50$ times  by the cD galaxy of
the cluster, 6$''$ apart.  Since this is the brightest LBG galaxy
known, it has been widely studied in the optical, submillimeter and
radio bands, both in imaging and spectroscopy.  Optical low resolution
spectra obtained with Keck/LIRIS (Pettini et al., 2000) showed the
complex structure of the interstellar medium of the galaxy.

Observations of cB58 were performed using the UV--visual Echelle
Spectrograph (UVES) of the ESO Very Large Telescope (VLT). This
instrument/telescope combination is the best possible today for
\lya~forest observations. The main motivation of the program was to
investigate the intervening absorption and compare its properties with
the equivalent in QSO forests. The results of the analysis of the
intergalactic \lya~forest distributed along the cB58 sight line are
presented in this paper.  The second goal was the study of the
metallicity and dust content in the interstellar medium of the galaxy
itself. These results will be presented in a future paper.

\section{OBSERVATIONS AND DATA REDUCTION}

The galaxy MS1512--cB58 ($\alpha=15^h 14^m 22.2^s$ $\delta=+36^\circ
36' 24''$, J2000) was observed with the high resolution UV--visual
echelle spectrograph (UVES, Dekker et al., 2000) mounted at the UT2
telescope of the VLT observatory, at Paranal (Chile). The program was
run in Service Observing mode between March and August 2000. The log
of the observations is shown in Table \ref{t1}. The red and blue arms
of UVES were used, centered at $\lambda=5200$ \AA~and 4370 \AA,
respectively.  Due to the high declination of the object with respect
to the latitude of the VLT observatory ($-24^\circ 40'$), the
observations were performed in chunks of $\sim1.5$ hours around the
meridian ($\sim 29^\circ$ above the horizon) to minimize atmospheric
absorption. Indeed, the airmass ranged between 2.1, the minimum value
for this object observed from Paranal, and 2.3. In order to maximize
the signal--to--noise ratio (S/N), the CCD was rebinned $2\times2$
pixels during observations.  Due to good seeing conditions, the slit
aperture was always set to $1''\times9''$ and $1''\times 10''$ for the
Red and Blue arm set--ups, respectively, and its inclination was along
the galaxy apparent major axis.

The data reduction has been performed using a MIDAS package
specifically implemented for UVES data. The final useful spectral
coverage was $\lambda\lambda=4150-5000$ \AA~for the blue arm, and
$\lambda\lambda=4500-5164$ \AA~and $\lambda\lambda=5241-6206$ \AA~for
the red arm. The pixel size was set to a constant value of 0.2 \AA~in
the final combined spectrum. The final resolution element
corresponding to two pixels then ranges from FWHM $=29$ \kms~at 4150
\AA, to 26 \kms~at the \lya, to 19 \kms~at 6206 \AA.  The noise
spectrum, used to determine the errors on the line parameters, has
been calculated through photon statistics propagation of object and
sky spectra and the detector read--out--noise.  The final S/N ratio
per resolution element is S/N$ \simeq3$ at $\lambda=4150$ \AA, S/N$
\simeq7$ at $\lambda=4550$ \AA, S/N$ \simeq8$ at $\lambda=5500$ \AA,
and S/N $\simeq10$ at $\lambda=6100$ \AA.

\begin{figure}
\caption[f1]{The Lyman--$\alpha$ forest of cB58 (thick line) with the
fitted Voigt profiles (thin smooth line). The spectrum of the noise per
resolution is also shown. Arrows indicate the position of metal 
and \lya~lines associated with the galaxy itself. Large and small
ticks mark \lya~clouds with HI column density $N_{\rm HI}\geq 10^{13.7}$ 
cm$^{-2}$ and $N_{\rm HI}<10^{13.7}$ cm$^{-2}$, respectively.}
\label{f1}
\centerline{{\epsfxsize=16.cm \epsfysize=21.cm \epsfbox{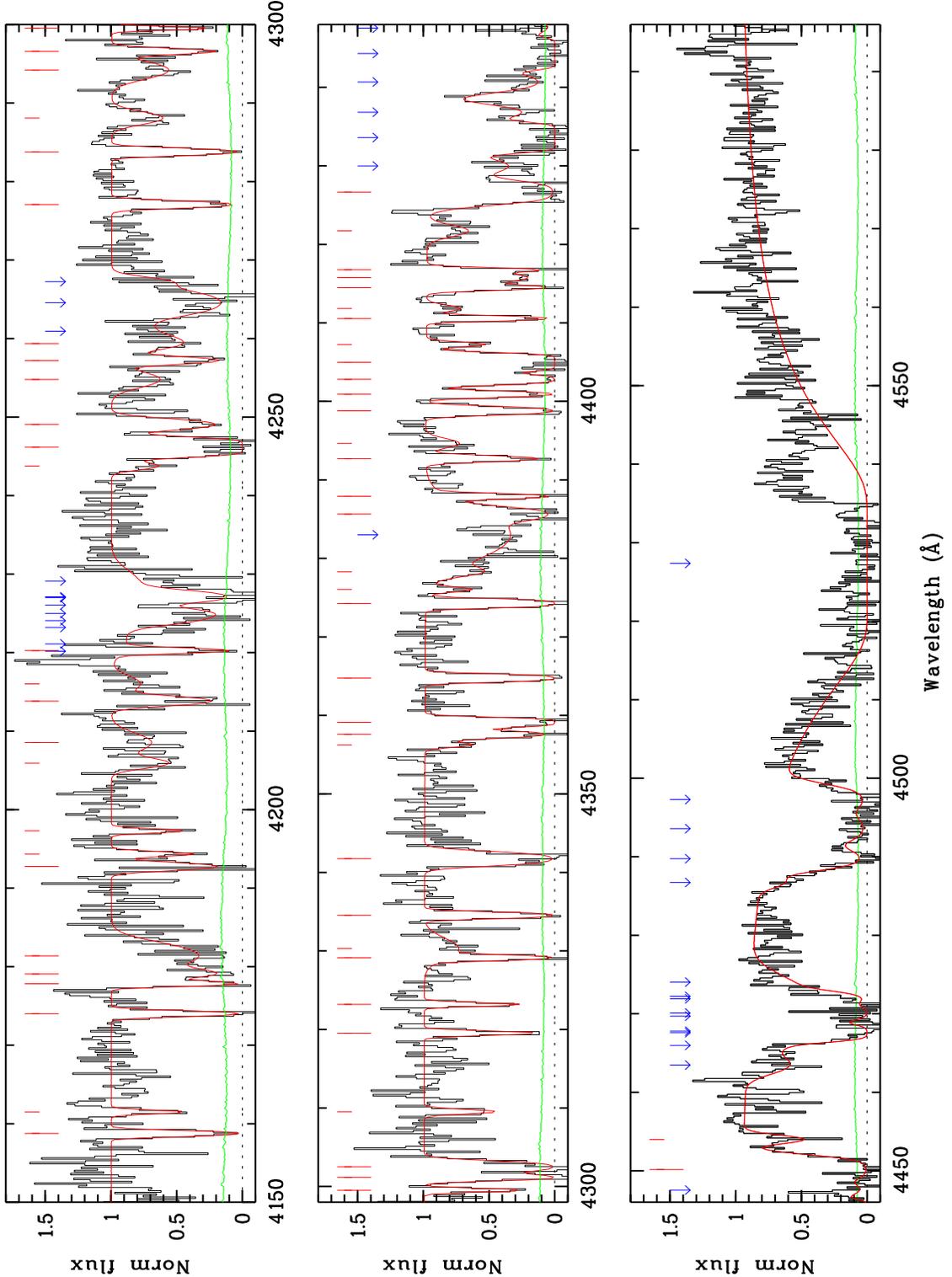}}}
\end{figure}

The fit to the continuum blueward of the galaxy \lya~was performed
interpolating between points of the observed flux free of apparent
absorption. Although this is not particularly easy due to the
non--excellent S/N ratio of the spectrum, at these redshifts the line
density of \lya~clouds allows the detection of several regions that
are most likely unabsorbed.  The final normalized galaxy and noise per
resolution element spectra in the interval $\lambda\lambda=4150-4595$
\AA~are shown in Fig.~\ref{f1}.

\section{THE LYMAN ALPHA FOREST OF cB58}

The magnification of cB58 and the excellent performance of UVES make
this galaxy the only suitable candidate today to study the
intergalactic matter distributed along a galaxy sight line, similarly
to what is normally done for QSO sight lines. The study of QSO
\lya~forests has led in the past to a broad understanding of the
formation and evolution of the Universe.  Very briefly we mention
successful experiments to test the intergalactic medium (IGM)
reionization through the Gunn--Peterson effect for HI and HeII, the
early metal pollution of the IGM, and the intensity of the UV
background radiation, as discussed by Songaila (1998), Songaila et
al. (1999), Heap et al. (2000), Ellison et al. (2000), just to mention
the most recent publications on these subjects.

For the first time we can test some of these results considering a
galaxy line of sight. Any deviation from the expected properties of
the IGM might be strongly connected to a deviation of the galaxy
surrounding environment with respect to the QSO ones.  To study the
cloud distribution in cB58, we have performed line fitting with Voigt
profiles using the MIDAS package FITLYMAN (Fontana \& Ballester,
1985).  The wavelength range considered here is
$\lambda\lambda=4150-4550$ \AA.  The upper limit is set by the
wavelength of the galaxy \lya~line.  Fig.~\ref{f1} shows the
\lya~forest of cB58, together with the fitted absorption profiles. In
this range some metal lines associated with cB58 itself have been
identified and fitted for completeness. No other metal lines
associated with intervening metal systems at $z=0.828$ and 1.339
reported by Pettini et al. (2000), and at $z=2.007$, 2.117 and 2.660
identified by us, have been found in this interval.

In Table~\ref{t2} we report absorption line parameters for the
\lya~lines only, namely the redshift $z$, column density $N_{\rm HI}$,
and the Doppler parameter $b$.

The intrinsic spectrum of the galaxy shows a very large \lya~
absorption that extends almost 100 \AA~in the range
$\lambda\lambda\simeq4480-4570$ \AA. This large trough has been
considered by Pettini et al. (2000) as a single interstellar cloud at
$z=2.7240$ with an HI column density of $N_{\rm HI} =
7.5\times10^{20}$ \cm2, assuming that the peak around 4540 \AA~is due
to \lya~emission of cB58 redshifted to $z=2.7326$ ($v=+690$
\kms). Making the same assumption we find a slightly lower HI content,
but still consistent with Pettini et al.'s result, with $N_{\rm HI} =
6^{+1.4}_{-2}\times10^{20}$ \cm2~and redshift $z=2.72417$. However we
also notice that in our spectrum the core of this \lya~absorption
($\lambda\lambda=4520-4535$ \AA) is not completely black, and a small
residual mean flux of about 1$\sigma$ above the zero level is present.
This might indicate that the interstellar clouds are not fully
covering the star forming regions of the galaxy and/or that resonant
scattering fills up part of the absorption.

Other metal lines associated with cB58 are listed in Table
\ref{t3}.  Due to the large HI absorption and metal lines associated
with the galaxy, we have considered in our analysis the \lya~forest
shortward of $\lambda=4430$ \AA, namely, the redshift range $2.4137
\leq z \leq 2.6441$ ($\Delta z = 0.23$). Since the galaxy intrinsic
brightness is much lower than a typical distant QSO, the classical
proximity effect, that in a QSO decreases the cloud density within
$\sim 8$ Mpc from the source due to the high QSO UV ionizing flux, is
in the case of cB58 negligible. Indeed, taking the observed magnitude
and magnification, the absolute magnitude of the galaxy is $M_B \sim
-22$, that is several hundreds times fainter than a high redshift QSO
typically used for proximity effect studies. For instance, if we
compare cB58 with a QSO 400 times brighter, the ionization region is
20 times smaller. This is even smaller if we consider that the
spectrum of a galaxy is much softer than that of a QSO and that the
intensity of UV ionizing spectrum is much weaker. We conclude that the
proximity distance for cB58 is confined to within 1 Mpc from the galaxy
center. The upper redshift limit of $z=2.6441$ corresponds to a
comoving distance from the galaxy of about 90 $h^{-1}_{65}$ Mpc (we
adopt throughout the paper a flat Universe with non-zero cosmological
constant: $\Omega_m=0.3$ and $\Omega_\lambda=0.7$). In the following
sections we discuss how an excess of \lya~clouds can be present if the
galaxy is embedded in a rich environment which might be typical for a
star forming high redshift galaxy.

To make a comparison with mean absorption in QSO sight lines, we have
considered two approaches: $i)$ we have measured the mean flux
depression blueward of the cB58 \lya~line ($D_A$) and $ii)$ we
calculated the number density of \lya~clouds.

\subsection{Mean flux depression in cB58}

The mean flux depression $D_A$ overcomes the problem of moderate
signal--to--noise ratio in the spectrum. It is defined by the equation:

\begin{equation}
D_A \equiv 1 - \left\langle f_{obs}/f_{int} \right\rangle 
\end{equation}

\noindent
where $f_{obs}$ and $f_{int}$ are the observed and intrinsic galaxy
fluxes, respectively, in the \lya~forest between the rest frame
\lya~and \lyb~wavelengths. This quantity is also expressed in terms of
the effective optical depth  in the same region:

\begin{equation}
D_A \equiv  1-  e^{-\tau_{\rm eff}} 
\end{equation}

\noindent
and has been measured in the past in QSO sight lines when the data
quality did not allow an accurate column density determination of
absorption lines (e.g, Giallongo \& Cristiani, 1990). More recently
Fardal et al. (1998) have computed $D_A$ using different models of
intergalactic absorption and compared results with observations. The
observed points are obtained not directly from the spectra, but from
line lists of Keck data and other lower S/N data.  Their Fig.~3 gives
$D_A\sim0.1$ at $z\sim2.25$ and $D_A\sim0.25$ at $z\sim2.8$. In an
extensive study of the mean HI opacity in QSOs, Kim et al. (2001a)
report $\tau_{\rm eff}$ values in 13 QSOs observed with VLT/UVES and
Keck/HIRES and give a best fit to the data in the interval $1.6 < z <
4.3$ as $\tau_{\rm eff} = (0.0030\pm0.0008) (1+z)^{3.43\pm0.17}$.

\begin{figure}
\caption[f2]{The Lyman--$\alpha$ forest of cB58 (upper spectrum)
compared with QSO sight lines of HS1946+7458, HE1347--2457 and 
GB1759+7539.}
\label{f2}
\centerline{{\epsfxsize=17cm \epsfysize=17cm \epsfbox{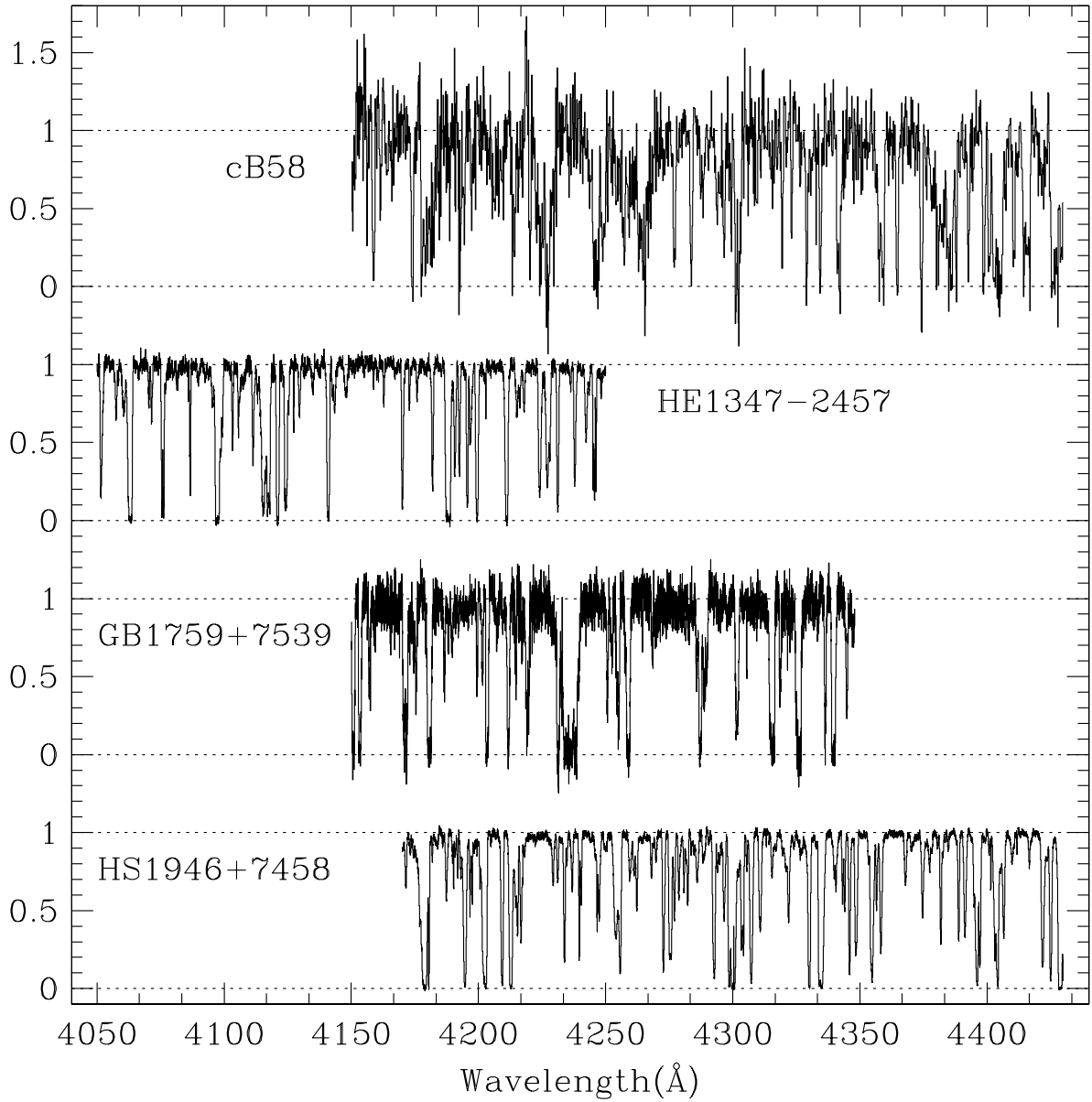}}}
\end{figure}

In our analysis, we considered the few very recent high resolution spectra
of QSO sight lines observed at similar redshifts, e.g. the Keck/HIRES
spectra of GB1759+7539 ($z_{em}=3.05$; Outram et al., 1999), and
HS1946+7658 ($z_{em}=3.051$; Kirkman \& Tytler, 1997), and the
VLT/UVES spectra of HE1347--2457 ($z_{em}=2.534$; Kim et al., 2001b)
and J2233--606 ($z_{em}=2.238$; Cristiani \& D'Odorico, 2001) obtained
during the science verification of the instrument.  Details of the QSO
spectra used are given in Table \ref{t4}, while in Fig.~\ref{f2} we
show the \lya~forest of cB58 and the comparison with QSOs in the region of
overlapping (except for J2233--606 that is too blue for cB58).  For
all the four QSO spectra, we have excluded the proximity effect
region.  For GB1759+7539, we have calculated the mean depression
redward of the QSO \lyb~emission and far from the proximity effect and
from a large HI absorption due to an intervening damped \lya~system at
$z=2.626$. For HE1347--2457 we have considered the region 4050
\AA~$<\lambda <$ 4250 \AA.  For J2233--606 we have excluded the region
with two strong HI absorptions ($N_{\rm HI}>10^{16}$ atoms \cm2)
associated with two metal systems at $z<1.943$.

\begin{figure}
\caption[f3]{Mean depression $D_A$ as a function of mean redshifts in
cB58 (open circle), and in different QSO lines of sight.  Triangles
are for J2233--606, HE1347--2457, GB1759+7539, and HS1946+7458, from
the lowest to the highest redshift, respectively. Squares are from
Keck/HIRES and VLT/UVES high resolution data (Kim et al., 2001a). The
solid and dashed curves are the best fit to $D_A$ $\pm1\sigma$,
respectively ($D_A=1-\exp(-0.003\times(1+z)^{3.43\pm0.17})$), obtained
by Kim et al. (2001a) in the redshift interval $1.6 < z <
4.3$.}
\label{f3} 
\centerline{{\epsfxsize=16cm
\epsfysize=16cm\epsfbox{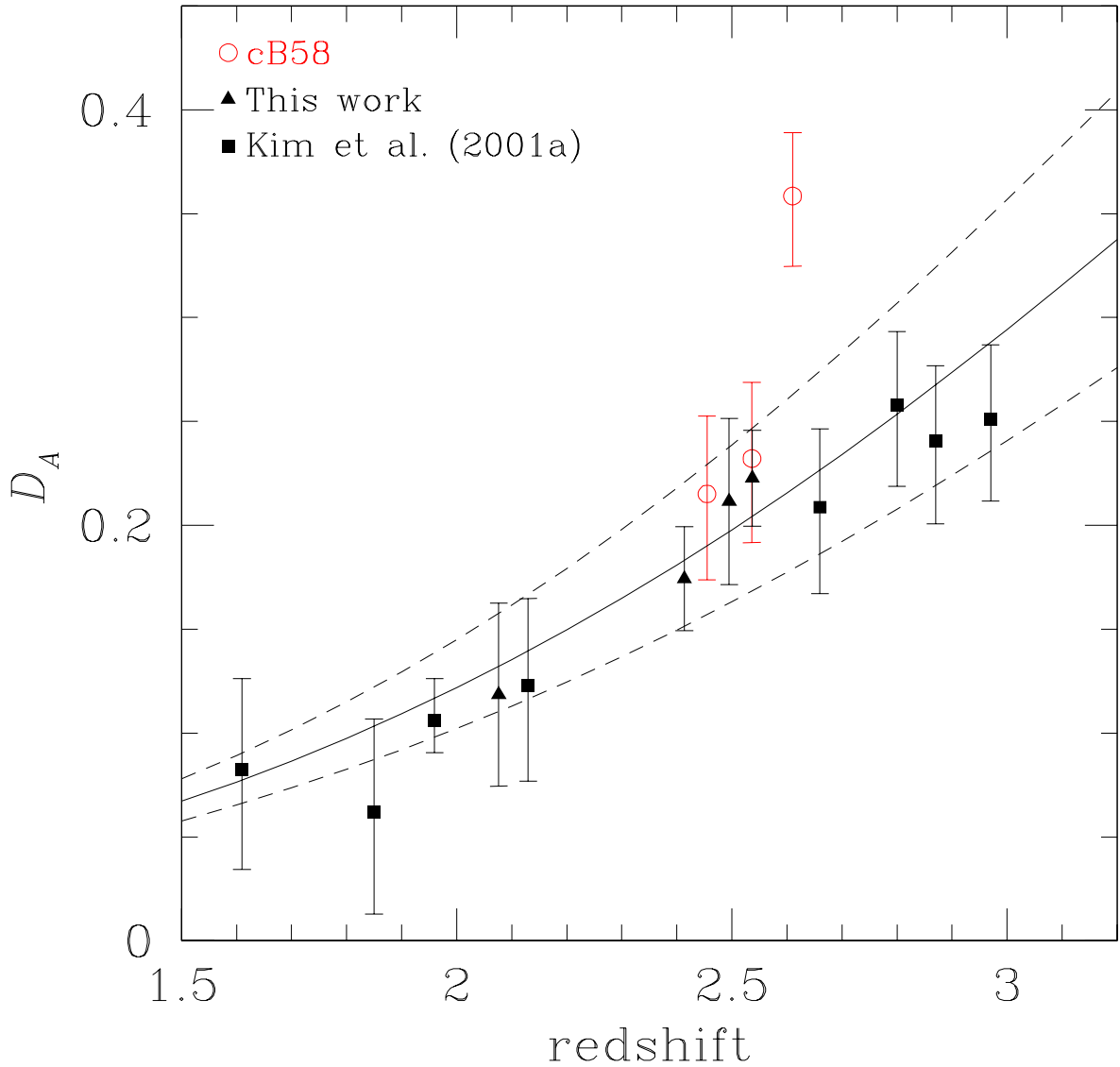}}}
\end{figure}

For cB58 we have calculate $D_A$ in the range 4150 \AA~$< \lambda <$
4430 \AA, corresponding to $2.4137 < z < 2.6441$.  For the
uncertainties on the mean depression, we have re--calculated $D_A$
considering as a new continuum the original estimate $\pm5$\% for
cB58, GB1759+7539 and J2233--606, and $\pm3$\% for HE1347--2457 and
HS1946+7658, respectively, according to S/N.  We take as our
error the uncertainty derived according to this procedure.  The value
of $D_A$ in the whole cB58 interval at $\langle z \rangle = 2.5289$ is
$D_A=0.263\pm0.037$. This is consistent with the value calculated from
the model of the fitted Voigt profiles listed in Table \ref{t2},
i.e. $D_A=0.255$. The total mean depression $D_A$ in the QSOs is given
in Table \ref{t4}, while in Table \ref{t5} we report $D_A$ in
different redshift intervals for cB58 and the QSOs, with the error
range between brackets. These values of $D_A$ have been combined with
results for other QSOs given by Kim et al. (2001a) and reported, as a
function of redshift, in Fig.~\ref{f3}.

Considering the uncertainty and the cosmic variance that we expect in
different lines of sight, we can conclude that the $D_A$ values for
cB58 are consistent with those for QSOs in the first two redshift
bins. However, we notice that the mean depression in cB58 is
appreciably larger than expected from QSOs in the last redshift
bin. In the last redshift bin of cB58 we have $D_A=0.36\pm0.03$ at
$\langle z \rangle = 2.6104$, while the best fit to QSO sight lines
gives $D_A=0.22\pm0.04$ at the same redshift (Kim et al., 2001a), a
$\sim 3\sigma$ effect.

\begin{figure}
\caption[f4]{Spectrum of cB58 (lower spectrum) compared with that of
the nearby dwarf starburst galaxy NGC1705 at redshift $v=+569$ \kms~
(upper spectrum, Heckman et al., 2001). Upper and lower x--axis scales
give the rest frame and observed wavelength range. Vertical ticks in
mark the ISM multiplets of NI(1134) and FeII(1144), and the stellar
lines SiIV(1122), SiIV(1128) and CIII(1176).  The non marked features
blushifted by $\sim 600$ \kms~in NGC1705 are absorption lines of the
Milky Way.}
\label{f4}
\centerline{{\epsfxsize=13cm \epsfysize=19cm 
\epsfbox{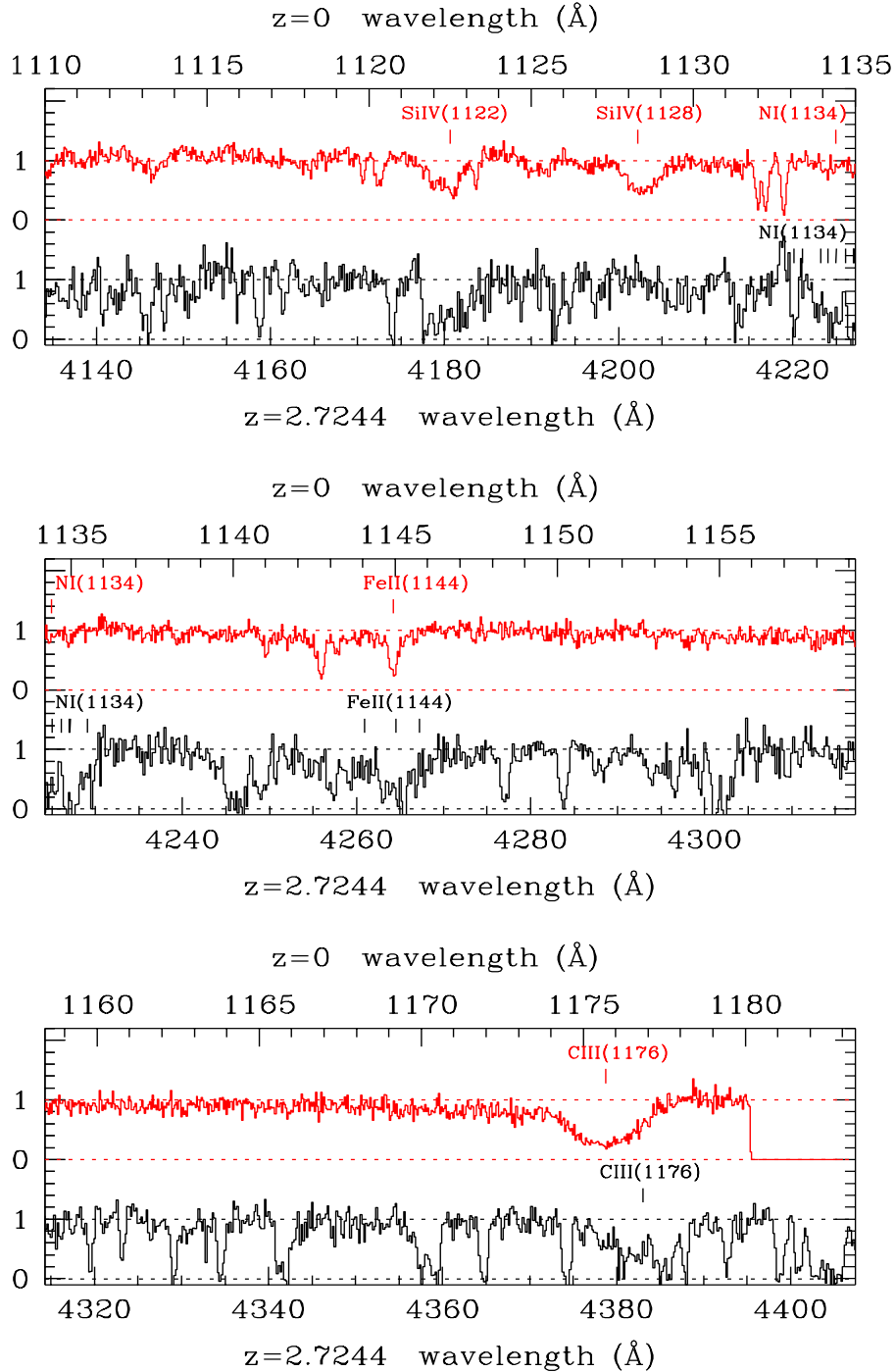}}}
\end{figure}

We notice that metal lines associated with cB58 could contribute to
$D_A$. This effect is marginally important in QSO sight lines, because
the powerful UV emission of QSOs highly ionize the foreground gas of
the host galaxy. To verify what this contribution is in cB58, we have
compared its UV spectrum with the UV spectrum of a nearby galaxy.
Thanks to the advent of FUSE, it is now possible to get high
resolution UV spectra of galaxies. One example is presented by Heckman
et al. (2001), who discuss the OVI outflow in the dwarf starburst
galaxy NGC1705. In Fig.~\ref{f4} we compare their FWHM $=30$
\kms~resolution spectrum with that of cB58 in the region of
overlapping. The rest frame overlapping range
$\lambda\lambda=1110-1183$ \AA~corresponds for cB58 to the observed
interval $\lambda\lambda=4134-4405$ \AA. The rest frame of cB58 is
calculated assuming a redshift of $z=2.7244$, that is the redshift of
the strongest central absorbing cloud in its interstellar medium
(ISM).  The absorption lines in NGC1705 show different contributions
coming from the ISM and high velocity clouds of the Milky Way, and the
ISM of NGC1705 itself, spanning a total velocity range of about 600
\kms~(Heckman et al., 2001). The strongest interstellar lines in
NGC1705, FeII$\lambda1144$ or NI$\lambda1134$, are also identified in
cB58. However, the involved column densities are about an order of
magnitudes higher in cB58. The strong absorption on the blue side of
the FeII complex, on the other hand, is due to the Galactic ISM, where
metallicity and column densities are much larger.

The strong SiIV absorption doublet at $\lambda\lambda=1122, 1128$
\AA~and the CIII absorption line at $\lambda=1176$ \AA~in the spectrum
NGC1705 are due to OB stars (Walborn \& Bohlin, 1996).  In cB58 the
absorption at $\lambda \sim 4383$ \AA~can be identified with CIII
absorption redshifted by about 300 \kms~with respect to the ISM.  The
fitted profile shown in Fig.~\ref{f1} has been obtained assuming a
width of 250 \kms~for this line, very similar to what observed for
NGC1705.  The contribution of the SiIV doublet in cB58 seems to be not
very significant (the first component is basically absent, while the
second one is very weak).  From unpublished HST/STIS high resolution
data of NGC1705 (Heckman, private communication) we have seen that
there are no other strong absorptions longward of the CIII feature in
NGC1705, up to $\lambda\simeq 1189$ \AA, corresponding to
$\lambda\sim4430$ \AA~in the cB58 observed frame, so other further
contributions to $D_A$ caused by the cB58 ISM can be considered
negligible.

We have then recalculated $D_A$ in the last wavelength bin of the cB58
forest and excluded CIII absorption around $\lambda\sim4383$ and
found $D_A=0.336\pm0.033$.  This is still significantly larger
than what found in QSOs and supports the idea that the excess of
absorption at $z\sim2.61$ (150 $h_{65}^{-1}$ Mpc comoving from the
galaxy) is not due to the galaxy itself but is associated with the
intervening intergalactic matter.

\subsection{Number density of \lya~clouds in cB58}

The next comparison between LBG and QSO forests is through the number
density per unit redshift of \lya~clouds with HI column density higher
than a threshold $N_{th}$. This eliminates the problem of the
metal lines associated with cB58. The number density per unit
redshift is parameterized as $dn/dz \propto (1+z)^\gamma$.  For a
non--evolving population in the standard Friedmann universe with null
cosmological constant, $\gamma=1$ and 0.5 for $q_o=0.0$ and 0.5,
respectively, whereas a non--evolving slope can be approximated by
$\gamma\sim0.7$ for a non-zero cosmological constant universe (Kim et
al., 2001a).  Recent work on QSO sight lines at different redshifts
has shown that strong \lya~clouds ($N_{\rm HI}>10^{14}$ atoms \cm2)
evolve at high redshift ($\gamma\sim2$ for $z>1.5$, Kim et al., 2001a)
followed by a flat distribution at redshifts $z<1.5$
($\gamma\simeq0.16$, HST key program, Weymann et al., 1998; Penton et
al., 2000) consistent with no evolution. Tracking the number density
evolution has implied a comparison of low resolution with high
resolution data. This complicates the interpretation; however, the
results are quite striking.

\begin{figure}
\caption[f5]{Number density evolution of \lya~clouds for HI column density
lines with $N_{\rm HI}\geq 10^{14}$ atoms cm$^{-2}$. The straight line is 
the best fit for QSO forests with $z\geq2.3$.}
\label{f5}  
\centerline{{\epsfxsize=16cm \epsfysize=16cm \epsfbox{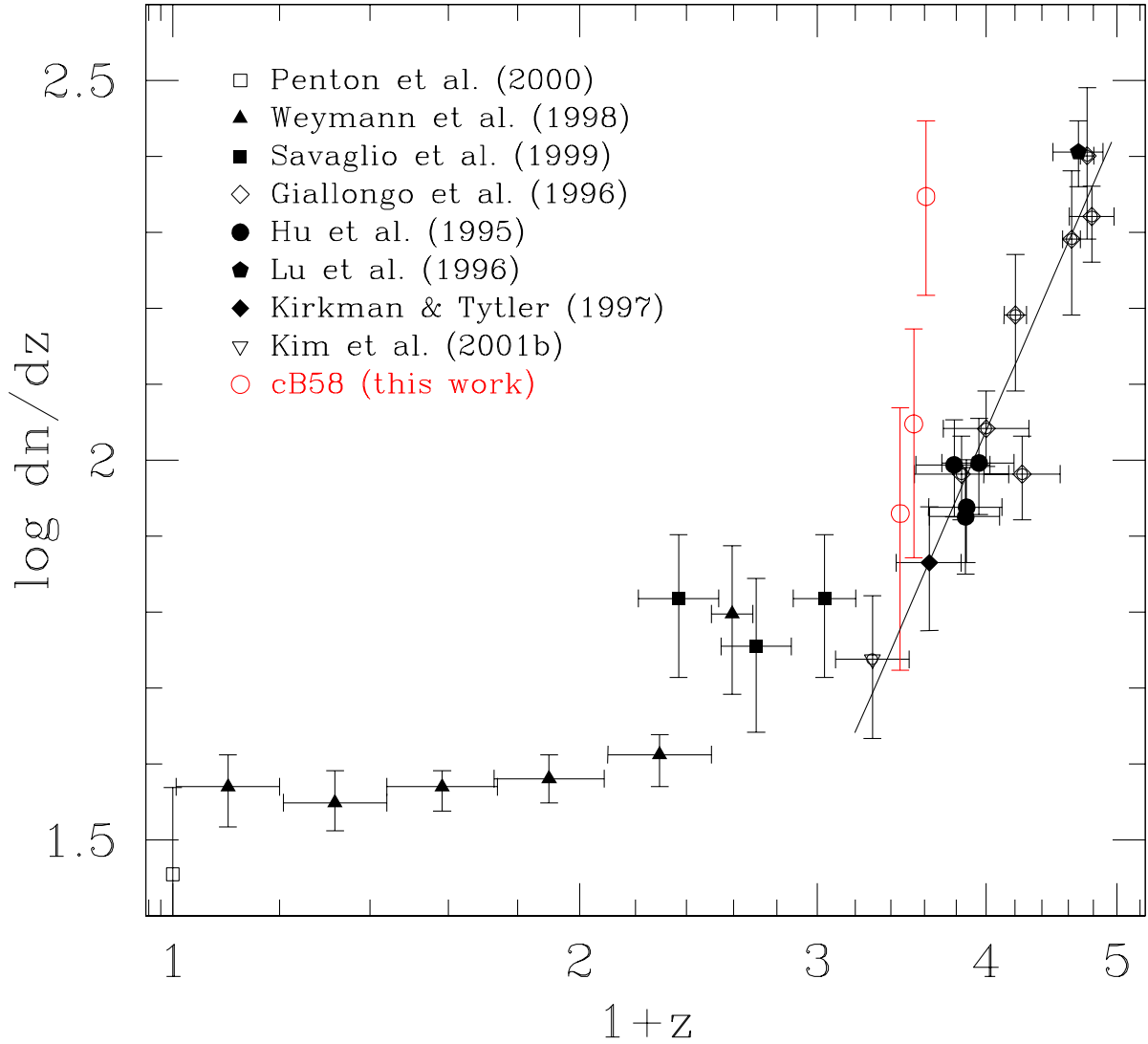}}}
\end{figure}

The number density evolution is different for different $N_{th}$,
being flatter at smaller $N_{th}$ for the large $z$ range (Penton et
al., 2000, Kim et al., 2001a).  In cB58 we have calculated $dn/dz$
dividing the redshift interval in three different redshift bins as
done in the previous section to calculate $D_A$. We also consider two
different $N_{th}$, namely, $10^{13.7}$ and $10^{14.0}$ atoms \cm2.
This is because the completeness of the \lya~sample varies along the
spectrum of cB58. Indeed, the $4\sigma$ completeness of HI column
densities obtained from \lya~lines with Doppler parameter in the range
20 \kms~$< b < 40$ \kms, is $N_{\rm HI} = 10^{13.5}, 10^{13.7}$, and
$10^{14.0}$ atoms \cm2~ for $\lambda>4270, 4180$, and 4150 \AA,
respectively. The limited resolution and S/N have the consequence of
underestimating the number of components observed in the absorption
profile and overestimating the HI column density. In other words, for
low or high $N_{th}$, the number of detected clouds is underestimated
or overestimated, respectively. In Tables \ref{t6} and \ref{t7} we
report the number of clouds with $\log N_{th}= 13.7$ and 14.0, while
in Fig.~\ref{f5} we show the number density evolution as a function of
redshift for clouds with $N_{th}\geq10^{14}$ \cm2.  When considering
results for $\log N_{th} = 14.0$ one should take into account that the
number of \lya~clouds might be overestimated if blending of lines is
important. However, this is unlikely to have a major effect on our
sight line, given the relative sparseness of \lya~lines at these
redshifts.  In the same tables, we also make a comparison with the
\lya~forests of GB1759+7539, HE1347--2457 and HS1946+7658.  Between
brackets we give the 1$\sigma$ range obtained using Poisson statistics
(Gehrels 1986). As typically done for QSO forests, the identified
metal lines in cB58 and in the QSOs associated with the galaxy itself
or other metal systems along the line of sight have been excluded from
the counting.

Tables \ref{t6} and \ref{t7} show that there is consistency in the
number of \lya~clouds between cB58 and QSO forests in the first two
redshift bins, whereas we find again a significant discrepancy when
the last redshift bin is considered.  This deviation appears more
clearly in Fig.~\ref{f5} for \lya~clouds with HI column density larger
than $10^{14}$ atoms \cm2 and is at the $2.7\sigma$
level with respect to the best fit to QSO forests in the interval $2 <
z < 4$.

As already found in the previous section where the mean depression is
considered, the number density of \lya~clouds gives a hint of an
excess of gas absorption in cB58 with respect to QSOs at
$z\simeq2.61$, or in the region within 150 $h_{65}^{-1}$ Mpc from the
galaxy. This excess appears to be more concentrated in the wavelength
range $\lambda\lambda=4370-4430$, at a mean redshift of $\langle z
\rangle =2.6194$. If this structure is isolated and not associated
with the galaxy itself, it is $\sim60~h^{-1}_{65}$ Mpc comoving wide
(velocity width of $\sim 4000$ \kms).

\section{DISCUSSION}

Observations of the \lya~forest have been fundamental in helping to
unveil the nature of the intergalactic gas. These studies have been
possible using large collecting areas pointed on very bright
sources. The only suitable candidates so far have been bright
($m_V<19$), distant ($z>2$) QSOs, observed with 4m and 10m class
ground based telescopes. Similar studies are still precluded if normal
Lyman Break Galaxies or radio galaxies are considered. This is because
no high-redshift galaxy brighter than $V \sim 21$ is known.  A unique
chance has been offered to us by VLT/UVES observations of the LBG
galaxy cB58. This galaxy at $z=2.724$ is gravitationally magnified
$\sim 50$ times by an intervening galaxy cluster at $z=0.37$ and its
optical apparent magnitude is $m_V=20.64$. This is by far the
brightest LBG known.

We have for the first time explored the \lya~forest of a high redshift
galaxy and compared the observed properties with those shown by QSO
sight lines. We found in cB58 an excess of absorption at a distance
from cB58 ($z=2.724$) of about 150 $h_{65}^{-1}$ Mpc comoving
($z\simeq2.61$). This has been quantified by both measuring the mean
flux depression $D_A$ due to HI absorption and the number density of
\lya~clouds with $N_{\rm HI} > 10^{14}$ atoms \cm2. The deviation is
in both cases significant at the $\sim 3 \sigma$ level. The large
\lya~absorption and some metal lines associated with cB58 prevent us
to detect \lya~intervening clouds for redshift larger than $z=2.644$,
so we cannot exclude the possibility that this high density region is
more extended. We now have two possible scenarios, depending on
the size of the region associated with this excess. 

\medskip
1) The excess is due to the presence of a super cluster of \lya~clouds
not associated with cB58. This super cluster of lines appears to be
concentrated at $2.5947 < z < 2.6441$ ($\Delta z \sim0.05$),
corresponding to a region $\sim 60~h_{65}^{-1}$ Mpc comoving
wide. This is an interesting possibility in view of the fact that a
very large structure on scales $\Delta z \sim0.1$ or $\sim
100~h_{65}^{-1}$ Mpc has been found at $z\sim3.1$ in the LBG galaxy
distribution (Steidel et al., 1998). Steidel et al.~also suggest that
similar clustering is common in LBG surveys and that LBGs represent
the progenitors of local massive galaxies.

\medskip
2) The excess is due to high-density gas associated with the
environment of cB58. In this case, this high density region could
extend all the way down to the galaxy. This has interesting
implications for the proximity effect. The proximity effect is used to
measure the intensity of the UV background radiation, assuming that
the gas distribution of the IGM is independent of the distance from
QSOs. Our findings seem to undermine this assumption, which has also
been questioned by Pascarelle et al. (2001), using a completely
different approach. These authors have analyzed the clustering
properties of galaxies and QSO absorption line pairs at low redshifts
($z\lsim 1$) and found that galaxies and \lya~clouds cluster around
QSOs. This effect extends up to a velocity range from the QSO of
$\Delta v \sim 3000$ \kms. In cB58 the clustering would be extended
down to $z\sim2.61$ ($\Delta z \sim0.11$), corresponding to a velocity
range of $\sim10,000$ \kms.  We can imagine that a clustering effect
similar to that found by Pascarelle et al. (2001) may occur in star
forming Lyman break galaxies, such as cB58. Indeed, its rest frame
optical emission lines give an amplified star formation rate (SFR) of
620 M$_\odot$ per year (Teplitz et al., 2000), which becomes $\sim10$
M$_\odot$ per year if corrected by the amplification factor of $\sim
50$. If cB58 is a starburst galaxy, as suggested by its SFR, then a
large density of starforming galaxies is expected in its environment,
and therefore a large number of absorbing clouds as well, consistent
with our findings.  Further evidence of the peculiarity of cB58 is
given by the large velocity range shown by its components: the ISM
absorption ($z=2.7244$) and the emission lines ($z=2.7290$) are spread
over almost 1000 \kms.

If the excess found around cB58 is typical in collapsed objects, the
QSO proximity effect has been underestimated so far and therefore the
intensity of the UV background is overestimated. This, in turn, has
important implications for estimates of the contribution of the
Lyman--$\alpha$ forest to the barionic content of the Universe.

\medskip

To assess the general validity of our results in other LBG forests, it
is necessary to analyze other good quality LBG spectra. This relies on
the chance that other gravitationally lensed LBGs are discovered,
hopefully with a magnification factor larger than that of cB58.  An
alternative solution to study the intrinsic properties of high
redshift galaxies and the intervening IGM clouds can be provided by
GRB afterglow observations. GRB afterglows have been found up to
redshifts of 4.5 and their optical magnitude can be as bright as
15$^{th}$ magnitude. These are perfect candidates for high resolution
spectroscopy, provided that they are promptly targeted soon after the
burst.

\acknowledgments

The authors wish to thank Tim Heckman, Tea--Sun Kim, David Kirkman,
Phil Outram and David Tytler for providing spectra and \lya~forest
lists. We are also grateful to our referee, David Kirkman, for very
insightful comments, and Tim Heckman for helpful discussions.  SS
acknowledges the support of the ESO and STScI Visitor Programs. SS
would like to express her gratitude for warm hospitality at ESO
Garching where the data reduction of UVES spectra has been performed.
Special thanks to Vanessa Hill, Andrea Modigliani, Stefano Cristiani
and Sandro D'Odorico for helping with the UVES pipeline.  The Paranal
and UVES teams are acknowledged for their effective support with
observations.

\newpage 

\begin{table}
\caption[t1]{Journal of the observations} \label{t1} 
\begin{center} 
\begin{tabular}{ccccc} 
\hline\hline&&&&\\[-10pt] 
Date & setup & seeing & exp time & airmass \\
(day/month/year) &       & ($''$) & (s)      &         \\
\hline&&&&\\[-8pt]
13/03/00 & R5200 & 0.6 & 5000 & 2.1 \\  
03/04/00 & R5200 & 0.5 & 5650 & 2.1	\\
26/04/00 & R5200 & 0.6 & 5400 & 2.1	\\
08/06/00 & R5200 & 1.1 & 5650 & 2.15	\\
10/06/00 & R5200 & 0.4 & 5650 & 2.08	\\
01/07/00 & R5200 & 0.7 & 5650& 2.32 \\
[2pt]\hline
total red arm      &       &       &33000 & \\
\hline\hline&&&&\\[-10pt]  
02/07/00 & B4370 & 1.3 & 5400 & 2.08 \\
04/07/00 & B4370 & 0.6 & 5400 & 2.24 \\
04/07/00 & B4370 & 0.6 & 5400 & 2.09 \\
05/07/00 & B4370 & 0.4 & 5400 & 2.32 \\
05/07/00 & B4370 & 0.4 & 5400 & 2.07 \\
26/07/00 & B4370 & 0.7 & 5400 & 2.07 \\
28/07/00 & B4370 & 1.5 & 5400 & 2.07 \\
02/08/00 & B4370 & 0.7 & 5400 & 2.07 \\
[2pt]\hline
total blue arm &       &           &43200 & \\
[2pt]\hline
\end{tabular}
\end{center}
\end{table}

\begin{table}
\caption[t2]{Line parameters of the \lya~forest of cB58} \label{t2} \begin{center} 
\begin{tabular}{ccccl} 
\hline\hline&&&& \\[-10pt] 
\#   & $\lambda_{obs}$ & $z$ & $\log N_{\rm HI}$ & $b$ \\ 
     & (\AA)       &     & (\cm2)            & (\kms) \\ 
\hline&&&&  \\[-8pt] 
1 & $4158.75\pm0.05$ & $2.42095\pm0.00004$ & $14.90\pm2.61$ & $15\pm13$ \\
2 & $4161.48\pm0.07$ & $2.42320\pm0.00006$ & $13.46\pm0.19$ & $14\pm9$ \\
3 & $4174.00\pm0.05$ & $2.43350\pm0.00004$ & $14.38\pm0.84$ & $19\pm10$ \\
4 & $4177.81\pm0.08$ & $2.43663\pm0.00007$ & $14.83\pm2.47$ & $14\pm10$ \\
5 & $4179.07\pm0.30$ & $2.43767\pm0.00025$ & $13.99\pm0.35$ & $50\pm47$ \\
6 & $4181.38\pm0.64$ & $2.43957\pm0.00053$ & $14.29\pm0.17$ & $132\pm41$ \\
7 & $4192.76\pm0.08$ & $2.44893\pm0.00007$ & $13.96\pm0.12$ & $37\pm8$ \\
8 & $4194.33\pm0.09$ & $2.45022\pm0.00007$ & $13.60\pm0.14$ & $23\pm9$ \\
9 & $4197.30\pm0.08$ & $2.45266\pm0.00007$ & $13.47\pm0.16$ & $17\pm11$ \\
10 & $4205.92\pm0.19$ & $2.45975\pm0.00016$ & $13.64\pm0.14$ & $57\pm19$ \\
11 & $4208.54\pm0.40$ & $2.46191\pm0.00033$ & $13.73\pm0.12$ & $110\pm37$ \\
12 & $4213.82\pm0.10$ & $2.46626\pm0.00008$ & $13.97\pm0.10$ & $46\pm8$ \\
13 & $4216.01\pm0.52$ & $2.46805\pm0.00043$ & $13.51\pm0.23$ & $97\pm67$ \\
14 & $4220.25\pm0.06$ & $2.47155\pm0.00005$ & $14.00\pm0.38$ & $20\pm9$ \\
15 & $4243.77\pm0.13$ & $2.49089\pm0.00011$ & $13.43\pm0.10$ & $42\pm13$ \\
16 & $4246.17\pm0.04$ & $2.49286\pm0.00003$ & $15.05\pm0.49$ & $54\pm12$ \\
17 & $4249.05\pm0.07$ & $2.49524\pm0.00006$ & $14.06\pm0.06$ & $52\pm6$ \\
18 & $4254.79\pm0.23$ & $2.49996\pm0.00019$ & $13.76\pm0.10$ & $92\pm28$ \\
19 & $4257.21\pm0.07$ & $2.50195\pm0.00006$ & $13.95\pm0.10$ & $35\pm7$ \\
20 & $4259.37\pm0.17$ & $2.50372\pm0.00014$ & $13.87\pm0.08$ & $80\pm20$ \\
21 & $4277.07\pm0.04$ & $2.51828\pm0.00003$ & $14.08\pm0.14$ & $28\pm5$ \\
22 & $4283.77\pm0.03$ & $2.52379\pm0.00002$ & $14.18\pm0.19$ & $26\pm5$ \\
23 & $4288.09\pm0.14$ & $2.52734\pm0.00012$ & $13.68\pm0.07$ & $71\pm13$ \\
24 & $4294.19\pm0.22$ & $2.53236\pm0.00018$ & $13.88\pm0.10$ & $98\pm31$ \\
25 & $4296.58\pm0.06$ & $2.53433\pm0.00005$ & $13.82\pm0.11$ & $26\pm7$ \\
26 & $4299.52\pm0.06$ & $2.53675\pm0.00005$ & $13.82\pm0.56$ & $11\pm6$ \\
27 & $4301.15\pm0.08$ & $2.53809\pm0.00007$ & $14.71\pm1.88$ & $18\pm13$ \\
28 & $4302.50\pm0.13$ & $2.53920\pm0.00011$ & $14.40\pm0.18$ & $40\pm10$ \\
29 & $4309.53\pm0.06$ & $2.54498\pm0.00005$ & $13.47\pm0.11$ & $19\pm5$ \\
30 & $4319.53\pm0.02$ & $2.55321\pm0.00002$ & $14.27\pm2.07$ & $12\pm12$ \\
[2pt]\hline
\end{tabular}
\end{center}
\end{table}

\begin{table}
\setcounter{table}{1}
\caption[t2]{\it -- Continued} \label{t2} 
\begin{center} 
\begin{tabular}{ccccl} 
\hline\hline&&&& \\[-10pt] 
\# & $\lambda_{obs}$ & $z$ & $\log N_{\rm HI}$ & $b$ \\ 
   & (\AA)       &     & (\cm2)            & (\kms) \\ 
\hline&&&&  \\[-8pt] 
31 & $4323.20\pm0.04$ & $2.55623\pm0.00003$ & $14.16\pm1.73$ & $8\pm7$ \\
32 & $4329.14\pm0.43$ & $2.56112\pm0.00035$ & $15.50\pm2.06$ & $10\pm11$ \\
33 & $4330.32\pm0.47$ & $2.56209\pm0.00039$ & $13.68\pm0.12$ & $119\pm41$ \\
34 & $4334.54\pm0.04$ & $2.56556\pm0.00003$ & $15.17\pm2.61$ & $15\pm9$ \\
35 & $4341.74\pm0.05$ & $2.57148\pm0.00004$ & $14.42\pm0.13$ & $49\pm6$ \\
36 & $4356.25\pm0.10$ & $2.58342\pm0.00008$ & $13.26\pm0.12$ & $24\pm12$ \\
37 & $4357.59\pm0.05$ & $2.58451\pm0.00004$ & $14.11\pm0.10$ & $31\pm6$ \\
38 & $4359.11\pm0.04$ & $2.58577\pm0.00003$ & $15.66\pm2.68$ & $20\pm12$ \\
39 & $4364.74\pm0.02$ & $2.59040\pm0.00002$ & $15.83\pm1.77$ & $15\pm6$ \\
40 & $4374.26\pm0.03$ & $2.59823\pm0.00002$ & $15.97\pm1.80$ & $17\pm6$ \\
41 & $4376.04\pm0.12$ & $2.59969\pm0.00010$ & $13.22\pm0.15$ & $24\pm12$ \\
42 & $4378.29\pm0.20$ & $2.60154\pm0.00016$ & $13.45\pm0.15$ & $50\pm20$ \\
43 & $4385.65\pm0.11$ & $2.60760\pm0.00009$ & $14.23\pm0.15$ & $53\pm10$ \\
44 & $4387.92\pm0.05$ & $2.60946\pm0.00004$ & $14.57\pm1.97$ & $13\pm13$ \\
45 & $4392.70\pm0.04$ & $2.61340\pm0.00003$ & $13.93\pm0.11$ & $24\pm4$ \\
46 & $4394.66\pm0.46$ & $2.61501\pm0.00038$ & $13.34\pm0.28$ & $52\pm36$ \\
47 & $4398.80\pm0.03$ & $2.61841\pm0.00002$ & $14.32\pm0.20$ & $28\pm4$ \\
48 & $4400.91\pm0.03$ & $2.62015\pm0.00002$ & $14.44\pm0.96$ & $19\pm9$ \\
49 & $4402.81\pm0.06$ & $2.62171\pm0.00005$ & $14.51\pm0.31$ & $35\pm10$ \\
50 & $4404.96\pm0.06$ & $2.62349\pm0.00005$ & $15.45\pm0.89$ & $42\pm13$ \\
51 & $4407.25\pm0.07$ & $2.62536\pm0.00006$ & $13.37\pm0.09$ & $23\pm9$ \\
52 & $4410.55\pm0.02$ & $2.62808\pm0.00002$ & $15.09\pm2.03$ & $10\pm7$ \\
53 & $4411.84\pm0.14$ & $2.62914\pm0.00012$ & $13.20\pm0.12$ & $36\pm12$ \\
54 & $4414.48\pm0.07$ & $2.63131\pm0.00006$ & $14.05\pm0.28$ & $19\pm8$ \\
55 & $4415.76\pm0.20$ & $2.63236\pm0.00016$ & $14.05\pm0.18$ & $52\pm28$ \\
56 & $4416.76\pm0.17$ & $2.63319\pm0.00014$ & $14.45\pm5.72$ & $10\pm13$ \\
57 & $4421.71\pm0.21$ & $2.63726\pm0.00017$ & $13.53\pm0.12$ & $64\pm24$ \\
58 & $4426.66\pm0.09$ & $2.64133\pm0.00007$ & $14.68\pm0.14$ & $86\pm16$ \\
59 & $4450.17\pm0.14$ & $2.66067\pm0.00012$ & $14.76\pm0.11$ & $93\pm10$ \\
60 & $4453.99\pm0.08$ & $2.66381\pm0.00007$ & $13.62\pm0.07$ & $44\pm8$ \\
[2pt]\hline
\end{tabular}
\end{center}
\end{table}

\begin{table}
\caption[t3]{Absorption metal lines associated with cB58} \label{t3} 
\begin{center} 
\begin{tabular}{clllc} 
\hline\hline&&&&\\[-10pt] 
\# & ID & $\lambda_{res}$ &  $\lambda_{obs}$ & $z$ \\
   &    & (\AA)         & (\AA)            & \\
\hline&&&\\[-8pt]
1 & NI & 1134.165 & 4220.18 & 2.72095 \\
2 & NI & 1134.415 & 4221.10 & 2.72095 \\
3 & NI & 1134.980 & 4223.21 & 2.72095 \\
4 & NI & 1134.165 & 4224.07 & 2.72438 \\
5 & NI & 1134.415 & 4225.00 & 2.72438 \\
6 & NI & 1134.165 & 4226.07 & 2.72615 \\
7 & NI & 1134.415 & 4227.00 & 2.72615 \\
8 & NI & 1134.980 & 4227.10 & 2.72438 \\
9 & NI & 1134.980 & 4229.11 & 2.72615 \\
10 & FeII & 1144.938 & 4260.94 & 2.72155 \\
11 & FeII & 1144.938 & 4264.56 & 2.72471 \\
12 & FeII & 1144.938 & 4267.25 & 2.72706 \\
13 & CIII & 1175.7 & 4383.00 & 2.60542 \\
14 & SiII & 1190.416 & 4429.98 & 2.72137 \\
15 & SiII & 1190.416 & 4433.60 & 2.72441 \\
16 & SIiI & 1190.416 & 4436.83 & 2.72713 \\
17 & SiII & 1193.290 & 4440.67 & 2.72137 \\
18 & SiII & 1193.290 & 4444.30 & 2.72441 \\
19 & SiII & 1193.290 & 4447.54 & 2.72713 \\
20 & NI & 1199.550 & 4463.47 & 2.72095 \\
21 & NI & 1200.223 & 4465.97 & 2.72095 \\
22 & NI & 1199.550 & 4467.58 & 2.72438 \\
23 & NI & 1200.710 & 4467.78 & 2.72095 \\
24 & NI & 1199.550 & 4469.70 & 2.72615 \\
25 & NI & 1200.223 & 4470.09 & 2.72438 \\
26 & NI & 1200.710 & 4471.90 & 2.72438 \\
27 & NI & 1200.223 & 4472.21 & 2.72615 \\
28 & NI & 1200.710 & 4474.02 & 2.72615 \\
29 & SiIII & 1206.500 & 4486.72 & 2.71879 \\
30 & SiIII & 1206.500 & 4489.75 & 2.72130 \\
31 & SiIII & 1206.500 & 4493.58 & 2.72448 \\
32 & SiIII & 1206.500 & 4497.28 & 2.72754 \\
33 & HI & 1215.670 & 4527.36 & 2.72417 \\
[2pt]\hline
\end{tabular}
\end{center}
\end{table}

\begin{table}
\caption[t4]{Mean depression $D_A$ in QSO sight lines}
\label{t4}
\begin{center} 
\begin{tabular}{cccccccc} 
\hline\hline&&&&&&&\\[-10pt] 
QSO & mag & $z_{em}$ & FWHM  & $\lambda$ range & $<z>$ & $\Delta z$ & $D_A$ \\
     &    &            &(\kms)   & (A)   &&& \\
\hline&&&&&&&\\[-8pt]
J2233--606$^a$ & $B17.5$ & 2.238  & 7  & 3580--3900 & 2.07649 & 0.2632 & $0.118\pm0.044$ \\ 
HE1347--2457$^b$& $R16.3$ & 2.534 & 7 & 4050--4250 & 2.41376 & 0.1645 & $0.174\pm0.025$ \\
GB1759+7539$^c$ & $R16.5$ & 3.05 & 7  & 4150--4348 & 2.49519 & 0.1629 & $0.212\pm0.040$ \\   
HS1946+7458$^d$ & $V15.9$ & 3.051 & 7.9 & 4170--4430 & 2.53714 & 0.2139 & $0.223\pm0.023$ \\
[2pt]\hline
\end{tabular}
\end{center}
$^a$ Cristiani \& D'Odorico, 2000; $^b$ Kim et al., 2001b; $^c$ Outram et al., 1999; $^d$ Kirkman \& Tytler, 1997.
\end{table}

\footnotesize
\begin{table}
\caption[t5]{Mean depression in different intervals of cB58 and QSO
sight lines.}\label{t5}
\begin{center} 
\begin{tabular}{cccccc} 
\hline\hline&&&&&\\[-10pt] 
       && \multicolumn{4}{c}{$D_A$} \\
\cline{3-6}\\
$\lambda$ range & $<z>$  & cB58 & GB1759+7539 & HE1347--2457 & HS1946+7458 \\
\hline&&&&&\\[-8pt]
4050--4150 & 2.3315 & --                   & --                   & 0.184 (0.159--0.208) & -- \\
4150--4250 & 2.4549 & 0.215 (0.174--0.253) & 0.234 (0.193--0.270) & 0.165 (0.138--0.189) & 0.219 (0.195--0.242)$^a$ \\
4250--4348 & 2.5363 & 0.232 (0.192--0.269) & 0.189 (0.146--0.228) & --                   & 0.240 (0.216--0.262) \\ 
4348--4430 & 2.6104 & 0.358 (0.325--0.389) & --                   & --                   & 0.206 (0.182--0.229) \\
[2pt]\hline
\end{tabular}
\end{center}
$^a$This value is calculated in the wavelength range 4170--4250 \AA, at $<z>=2.4631.$
\end{table}
\normalsize

\begin{table}
\caption[t6]{Number of \lya~clouds in cB58 and different QSO sight lines}\label{t6} 
\begin{center} 
\begin{tabular}{ccccccc} 
\hline\hline&&&&&&\\[-10pt] 
       &&& \multicolumn{4}{c}{Number of \lya~clouds $(\log N_{\rm HI} \geq 13.7$)} \\
\cline{4-7}\\[-10pt]
$\lambda$ range&$<z>$& $\Delta z$ & cB58 & GB1759+7539 & HE1347--2457 & HS1946+7658 \\
\hline&&&&&&\\[-10pt]
4050--4150 & 2.3315 & 0.0823 & --   & --   & 10 (6.9--14.3) & -- \\
4150--4250 & 2.4549 & 0.0823 & 11 (7.7--15.4) & 12 (8.6--16.6) & 11 (7.7--15.4) & 10 (6.9--14.3)$^a$ \\
4250--4348 & 2.5363 & 0.0806 & 15 (11.2--20.0) & 7 (4.4--10.8)  & --   & 12 (8.6--16.6) \\
4348--4430 & 2.6104 & 0.0675 & 16 (12.1--21.1) & --   & --   & 9 (6.1--13.1) \\ 
[2pt]\hline
\end{tabular}
\end{center}
$^a$This value is calculated rescaling from the value obtained in the wavelength range 4170--4250 \AA, at $<z>=2.4631$.
\end{table}

\begin{table}
\caption[t7]{Number of \lya~clouds in cB58 and different QSO sight lines}\label{t7}  
\begin{center} 
\begin{tabular}{ccccccc} 
\hline\hline&&&&&&\\[-10pt] 
       &&& \multicolumn{4}{c}{Number of \lya~clouds $(\log N_{\rm HI} \geq 14.0$)} \\
\cline{4-7}\\[-10pt]
$\lambda$ range&$<z>$& $\Delta z$ & cB58 & GB1759+7539 & HE1347--2457 & HS1946+7658 \\
\hline&&&&&&\\[-10pt]
4050--4150 & 2.3315 & 0.0823 & --   & --  & 8 (5.2--11.9) & -- \\         
4150--4250 & 2.4549 & 0.0823 & 7 (4.4--10.8) & 9 (6.1--13.1) & 5 (2.9--8.4) & 6 (3.6--9.6)$^a$ \\
4250--4348 & 2.5363 & 0.0806 & 9 (6.1--13.1) & 4 (2.1--7.2) & --  & 6 (3.6--9.6) \\
4348--4430 & 2.6104 & 0.0675 & 15 (11.2--20.0) & --  & --  & 2 (0.7--4.6) \\ 
[2pt]\hline
\end{tabular}
\end{center}
$^a$This value is calculated rescaling from the value obtained in the wavelength range 4170--4250 \AA, at $<z>=2.46311$.
\end{table}

\newpage 

\end{document}